%% file: cp_decomp_080526.tex
\documentclass[
    aps,prb,twocolumn,
	groupedaddress,superscriptaddress,
	amsfonts,amssymb,amsmath,
	citeautoscript,longbibliography,
	letterpaper
	]{revtex4-2}

\usepackage{import}                         %
\usepackage{graphicx}
\usepackage{epstopdf}
\usepackage{amsmath} 
\usepackage{bm}
\usepackage{amssymb}
\usepackage{quotes}
\usepackage{transparent}
\usepackage{dcolumn}
\usepackage{multirow}
\usepackage{cancel} 
\usepackage{mdframed}
\usepackage{dsfont}
\usepackage{slashed}
\usepackage[table]{xcolor}
\input{preamble}
\makeatletter
\@ifundefined{@tempdimc}{\newdimen\@tempdimc}{}
\def\CT@color{%
  \global\let\CT@do@color\CT@@do@color
  \color}
\def\CT@setup{%
  \@tempdimb4pt
  \@tempdimc4pt}
\makeatother

\begin{document}
\rmfamily
\raggedbottom

\title{Simulating multimode nonlinearities at very high mode count via parallel factor decomposition}
\author{Nicholas Rivera}
\affiliation{School of Applied and Engineering Physics, Cornell University. Ithaca, NY 14853.}

\begin{abstract}
A variety of problems at the frontier of nonlinear optics involve strongly multimode interactions. Simulating these dynamics is often computationally challenging due to the complexity of evaluating a large number of nonlinear overlaps. This is especially acute in ultrafast Kerr effect physics, where simulations in a modal representation have a generic complexity scaling quartically with the number of modes. This has limited the largest number of modes retainable in simulations in practice to a few tens of modes (often fewer), even when run on modern GPUs. Here, I show how by taking advantage of parallel factor decompositions of the nonlinear overlap tensor, a far larger number of modes can be accommodated. Realistic values for generic index profiles without special symmetries or separability can be in the thousands, and with special symmetries, far larger. I show that simulations with $>$1,000 modes can be done on modern hardware, illustrating this with the example of a multimode soliton propagating over a meter of GRIN fiber. I further show that the improved scaling immediately makes it practical to consider statistical effects, illustrating with the example of noise amplification in soliton fission. These results, implemented in public code with examples, should enable access to simulating the most demanding regimes of multimode nonlinear optics. Compression of the nonlinear interaction tensor may provide a route for more efficient simulation of nonlinear wave dynamics across a variety of other systems including acoustics, water waves, phononics, magnonics, and Bose condensates.
\end{abstract}

\maketitle

A variety of problems at the frontier of ultrafast nonlinear optics involve the Kerr effect in the highly multimode regime, which is known to host complex effects. Observations include beam-cleaning \cite{krupa2017spatial}, thermalization \cite{pourbeyram2022direct}, spatiotemporal modelocking \cite{wright2017spatiotemporal}, and geometric parametric instability \cite{wright2015controllable}. These phenomena, for the most part, have resisted general analytical treatments (particularly in the pulsed regime), and experimental probes of these systems are subject to considerable uncertainty in the initial complex spatiotemporal fields, and the exact spatial distribution of the index of refraction (which sets diffraction, dispersion, disorder, and nonlinearity). The complexity of multimode nonlinear optics (MMNLO \footnote{I will use MMNLO to mostly refer to third-order nonlinear physics but much of what is said here is relevant to other nonlinearities in highly multimode regimes. The treatment pursued here would work as well for multimode second-order nonlinear systems, thermo-optic systems, and other nonlinear wave systems.}) has motivated simplifying analytical approaches for understanding different regimes. Recent examples include attractor decomposition for spatiotemporal mode-locking \cite{wright2020mechanisms}, wave-turbulence theory \cite{picozzi2014optical} and optical thermodynamics  \cite{wu2019thermodynamic}. Nevertheless, simulations remain an indispensable route to validate theories, as well as design and explain experiments. 

Many phenomena in MMNLO can be understood through the multimode generalized nonlinear Schrodinger equation (MMGNLSE). To keep the scope of this paper bounded, I will consider a simpler form of the MMGNLSE: I will consider dispersion and instantaneous and delayed nonlinear response in the context of an isotropic medium such as amorphous silica. I will suppress polarization and disorder for now as it is not necessary to make the central point of the paper. I will also suppress the frequency dependence of the mode functions. I express the complex spatiotemporal electric field in a co-moving frame as $E(\boldsymbol{\rho},z,t) = e^{i\beta_0 z - i\omega_0 t}\sum\limits^M_{m=1} u_m(\boldsymbol{\rho})A_m(z,t)$, where $\boldsymbol{\rho},z$ are transverse and longitudinal spatial coordinates respectively, $t$ is a retarded time, and $m$ is a spatial mode index. The number of modes is $M$. In an isotropic medium, the corresponding MMGNLSE is expressed as:
\begin{widetext}
\begin{equation}
\label{eq:mmgnlse}
    \partial_z A_m  = \sum\limits_{k=0}^{\infty} \frac{i^{k+1}\beta_{m,k}}{k!} \partial_t^k A_m + i\left(1 + i\tau_S\partial_t \right)\sum\limits_{npq=1}^M \gamma_{mnpq} \left((1-f_R)A_n^*A_p + f_R[h_R \ast A_n^*A_p] \right) A_q.
\end{equation}
\end{widetext}
The main obstruction is the nonlinear contraction $\gamma_{mnpq}A_n^*A_pA_q$ (the Raman term is structurally similar). Since the number of non-zero elements (even after dropping ``small'' values) scales as $M^4$ where $M$ is the number of modes, the time to evaluate the contraction also scales as $M^4$. This, in practice, imposes a severe limitation on the number of modes which may be propagated, particularly for femtosecond pulses, where the complexity is further increased by the fact that this quartic term must be evaluated for each time or frequency point on one's grid. This has motivated the development of GPU-based massively-parallel algorithms for propagating the MMGNLSE \cite{wright2017multimode}. However, even these in practice have been limited to mode counts of a few tens, as even then, for modest lengths of propagation, simulation times exceed several hours (an example would be \cite{pourbeyram2022direct})

In this Letter, I show by explicit construction that the nonlinear overlap tensor governing Kerr nonlinear optics (the four-wave-mixing tensor) admits a low-rank tensor decomposition into \emph{parallel factors}, which strongly reduces the number of non-zero parameters which describe the tensor and strongly reduces the number of operations associated with the contraction. The ultimate complexity of the nonlinear contraction becomes approximately quadratic in general cases, which leads to orders-of-magnitude runtime reductions when the number of modes is large. I demonstrate this by examples with GRIN, step-index, and symmetry-breaking fibers to illustrate that the existence of a low-rank representation should hold broadly across different index profiles. I further explicitly demonstrate that this decomposition makes it practical to run multimode simulations with over a thousand modes over many nonlinear lengths, in the sub-hour timescale, as well as stochastic simulations which probe quantum statistical effects. These results should enable broader numerical exploration of some of the most demanding regimes of multimode nonlinear optics, and accelerate the design and validation of complex phenomena. A public code is available which contains the examples described in this Letter, as well as many others.

\begin{figure*}[!t]
    \centering
    \includegraphics[width=1\textwidth]{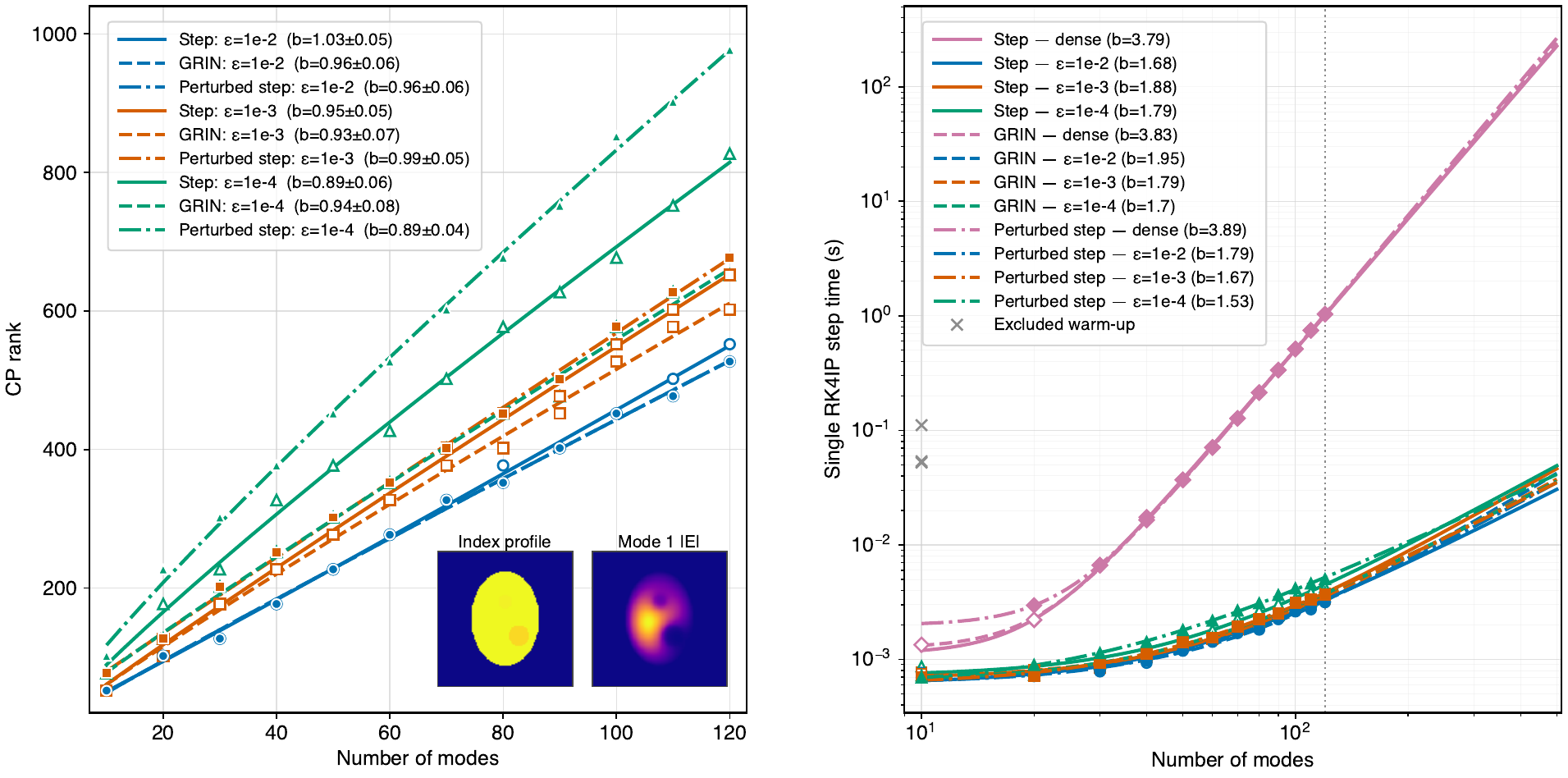}
    \caption{\textbf{Compression of the four-wave-mixing tensor.} CP rank and runtime as a function of mode count. CP rank leading to reconstruction of the four-wave mixing tensor at different error levels for a step-index fiber, a GRIN fiber, and a fiber with a perturbed elliptical index profile (index profile and fundamental mode shown in inset; the perturbations are regions of lower index). Fits of rank versus mode-count are of the form $R(M,\epsilon) \sim M^b$ showing a roughly linear scaling with rank in all cases. (b) Time to execute one propagation step of the MMGNLSE for the different systems, using the CP decomposed tensor and the standard dense approach (pink). Curves beyond 120 modes are an extrapolation. Fits are to the form $t(M,\epsilon) = c_0 + c_1 M^b$. }

    \vspace{12pt}
    \includegraphics[width=1\textwidth]{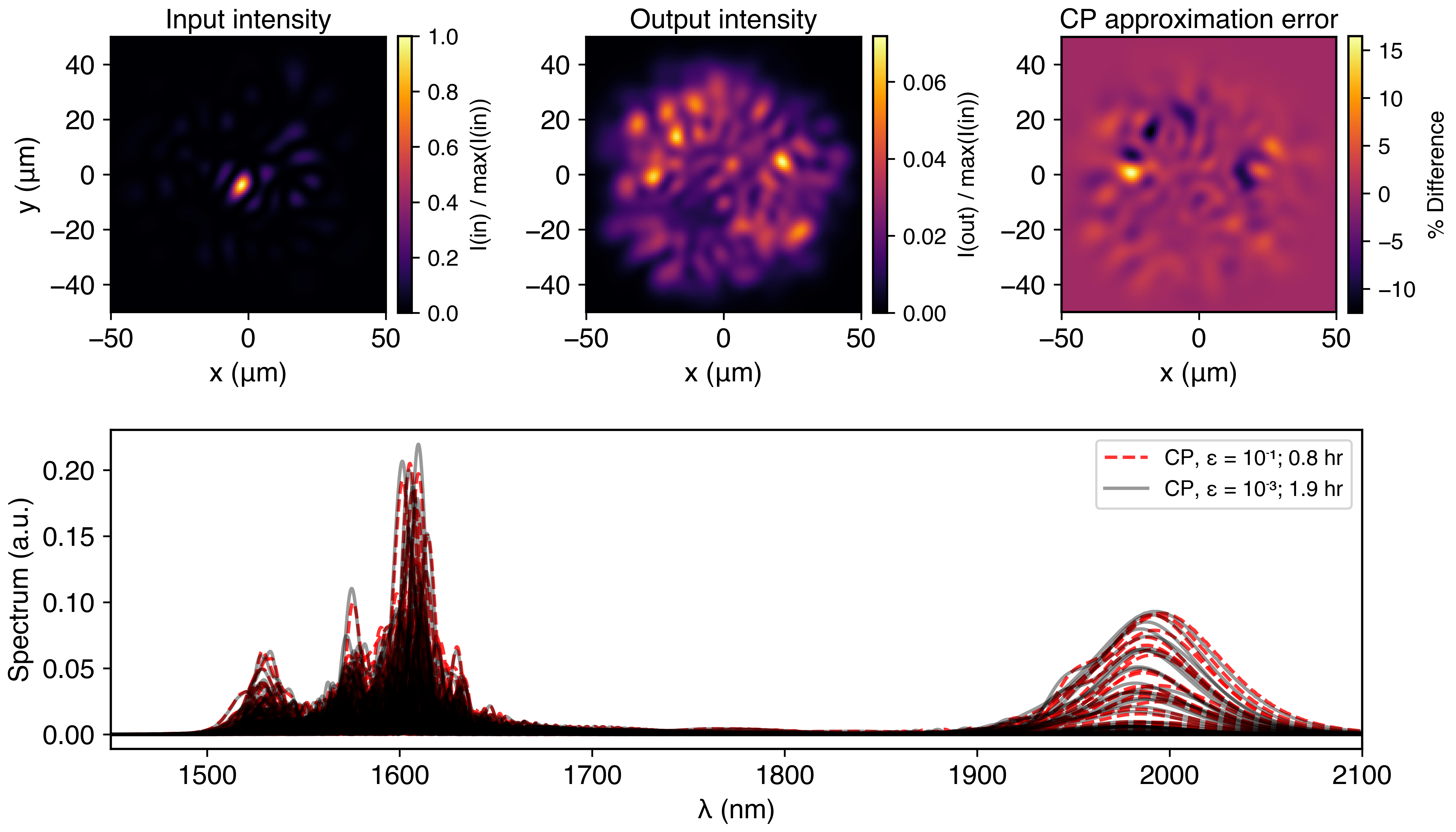}
    \caption{\textbf{Example of a multimode simulation involving 1,035 GRIN modes.}. An initial 120-mode soliton (with 1,035 total retained modes) launched in GRIN fiber. Spectrally-averaged intensity $I(x,y) \sim \sum_{\lambda} |E(x,y,\lambda)|^2|$ for the input (top left) and the output (top middle). Panel on the top right shows the difference in predicted intensities for CP approximations of $\epsilon = 10^{-1}$ and $10^{-3}$. (Bottom) Mode-resolved spectra for the two approximation errors. }
\end{figure*}

\section{Parallel factor decomposition of the four-wave mixing tensor} 
The main idea is to perform a canonical polyadic decomposition of the four-wave-mixing (4WM) tensor: $\gamma$ into parallel factors (a CANDECOMP/PARAFAC, or CP decomposition) \cite{kolda2009tensor}. This idea has recently been employed in a related context of compressing phonon-phonon interaction tensors, whose size grows greatly with the number of modes (momentum-space modes) \cite{luo2025tensor}.  The premise of CP decomposition is to seek an approximation of $\gamma$ of the form:
\begin{equation}
\label{eq:cp-decomposition}
    \gamma_{mnpq} \approx \hat{\gamma}_{mnpq} = \sum\limits_{r=1}^R \lambda_r u_{r,m}^{(1)}u_{r,n}^{(2)}u_{r,p}^{(3)}u_{r,q}^{(4)},
\end{equation}
which recasts for example the instantaneous Kerr term as a sum over $R$ dot products. I will call $R$ the rank in (loose) analogy to singular value decomposition. For this paragraph, I will focus only on the instantaneous Kerr term, but all figures shown include Kerr and Raman (as well as self-steepening). After replacing $\gamma$ by $\hat{\gamma}$, the Kerr contraction reads as: $\hat{\gamma}_{mnpq}A_n^*A_pA_q = \sum\limits_{r=1}^R \lambda_r u_{r,m}^{(1)}(u_{r}^{(2)}\cdot A^*)(u_{r}^{(3)} \cdot A)(u_{r}^{(4)} \cdot A)$. The delayed Raman contribution is also simplified greatly and does not require a separate factorization as $\gamma_{mnpq}$ appears in both Kerr and Raman terms in Eq.~\eqref{eq:mmgnlse}. If a sufficiently small $R$ suffices for $\hat{\gamma}$ to well-represent $\gamma$, then the number of elements of $\hat{\gamma}$ can be much less than $\gamma$, the tensor is compressed, and the contraction becomes cheaper to evaluate. More quantitatively: the number of quantities which parameterize $\hat{\gamma}$ is $(4M + 1)R$. The approximate runtime cost of the contraction, for all indices $m$ should naively also scale as $(4M+1)R$. The value of $R$ is determined by the reconstruction accuracy we want. In this work, I quantify the tensor reconstruction error $\epsilon$ (henceforth tensor error) in the simplest way possible, using the Frobenius norm: $\epsilon = ||\gamma-\hat{\gamma}||_F/||\gamma||_F$. The rank $R$ to achieve a tensor reconstruction error is a function of both $M$ and $\epsilon$: $R = R(M,\epsilon)$. It is natural that a smaller error will generically require a larger rank. It is also not surprising that more modes needs larger rank, since with more modes, one is trying to compress a higher-dimensional object. In order to assess whether the CP decomposition buys us a reduced complexity for the nonlinear contraction, I illustrate explicitly the Kerr term (the Raman-term complexity is the same up to a prefactor). The contraction involves $(4M+1)R(M,\epsilon)$ operations, and so a ``win'' requires $R(M,\epsilon) \ll M^3$. 

Figure 1a illustrates characteristic $R(M,\epsilon)$ curves for three types of nonlinear waveguides (all fiber, but this applies equally well for modal simulations of generic waveguides): graded-index, step-index, and a fiber with a perturbed elliptical profile. As can be seen, all three have very similar scaling, with a nearly \emph{linear} scaling of $R(M,\epsilon)$. That implies that the contraction should be quadratically better in runtime. Figure 1b, which shows the runtime scaling of a single $z$-step of the MMGNLSE, shows this for the CP-represented case of all three fibers (at different reconstruction error levels), and the original ("dense") representation. These runs were performed on an NVIDIA H100 SXM GPU: even then, on the scale of 100 modes, a typical $z$-step of the MMGNLSE takes on the order of seconds. The CP approach, at the lowest error, takes roughly 3 ms to execute a single step.

As an example of what this enables, in Figure 2, I show an end-to-end simulation which retains 1,035 modes (the first 45 mode groups of a GRIN fiber). The launch condition is a 0.2 $\mu$J, 120-mode soliton (250 fs FWHM pulse width, equal coefficients in the first 15 mode groups) propagating through one meter of GRIN fiber with core radius 100 $\mu$m microns and on-axis NA = 0.275. Polarization is omitted. The number of time points retained is $2^{13}$. The step size is 100 microns, corresponding to 10,000 propagation steps. At this mode count, the dense simulation is no longer reasonable to run. An extrapolation of Figure 1 (which is not perfectly justified due to the slightly different parameters) gives a time for a \emph{single} propagation step on the order of 1.5 hours on the same hardware, and a full propagation runtime of 1.75 years (Figure 1 considers $2^{13}$ time points).  With tensor compression at errors of $\epsilon = 10^{-1}, 10^{-3}$, the simulations are complete in roughly 0.8 and 1.9 hours (respectively), yielding an estimated speedup on the order of $10^4$. This order-of-magnitude expected from naive extrapolation of Figure 1 to 1,035 modes using a quadratic-in-modes scaling assumption for the speedup, even though the systems of Figure 1 are different from the one shown here. The modal spectra for the first 120 modes is shown at these two tensor errors: already the high-error representation gives a good approximation to the modal spectra. The spatial intensity errors between $\epsilon = 10^{-1}, 10^{-3}$ are also relatively small, and certainly small compared to typical experimental uncertainties in these types of systems. 

The timing reductions should allow for wider and deeper numerical exploration of complex effects associated with highly multimode nonlinear propagation. Statistical effects should get a special mention as they typically require running an ensemble of simulations, where one would ideally want on the order of a hundred or so trajectories for some reasonable convergence. Figure 3 demonstrates this in the context of shot-to-shot fluctuations in soliton fission dynamics, where we consider an initial twenty-mode soliton (105 modes retained in the simulation) with 0.1\% relative pulse energy fluctuations (treated simply as white noise, roughly 60 dB in excess of vacuum fluctuations) going through one meter of GRIN fiber. 
\clearpage
The top panel illustrates the initial time profile for the lowest mode while the bottom shows 120 shots of the time-profile for the highest intensity soliton that has broken off from the pulse. This run, which was executed serially on an H100 SXM, with 20,000 steps per run, took six hours. The average temporal intensity is shown in solid lines and we see that at 0.1\% relative pulse energy fluctuation, the soliton shot-to-shot fluctuations possess a large relative timing jitter that leads to a small and relatively smooth average temporal intensity. Compared to the single-mode case, it is observed that there is anticorrelation in the peak intensities between the peak heights of the first two modes in the soliton. Interestingly, despite this jitter, the twenty mode soliton remains temporally locked. The result clearly illustrates a strong coupling of noise to the average field, illustrating the importance of taking noise into account for measurements that average over many pulses. For systems like this, the noise is a necessary part of the input specification, and parallel factor decomposition techniques make it possible to address noise-driven physics in such systems.

\begin{figure}[t]
    \centering
    \includegraphics[width=0.5\textwidth]{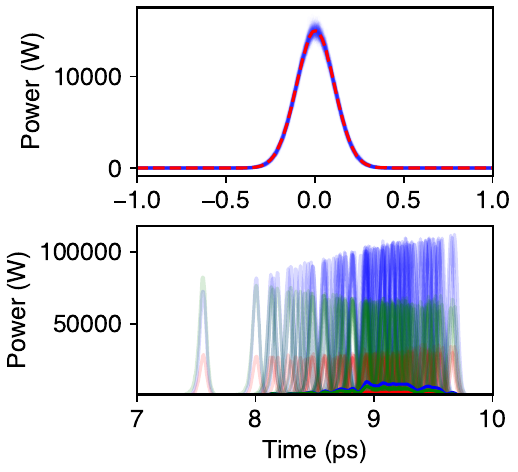}
    \caption{\textbf{Noise response in multimode soliton fission.} (a) Time-profile of initial pulse in lowest mode for 120 shots with a pulse energy fluctuation of 0.1\%. (b) Time profile of the highest-peak-power soliton that broke off of the initial pulse, shown for modes 1, 2, and 6, as well as the average time profile over 120 shots. 105 modes are retained in the simulation, which involves an initial twenty-mode (modes 1-20) launch propagating through one meter of fiber.}
\end{figure}

\section{Discussion}

\textit{When to use CP versus specialized methods}. I have illustrated a generic approach (CP decomposition) to get high-fidelity strong compression of the four-wave-mixing tensor, whose native complexity has strongly limited what multimode dynamics can be simulated. It relies on the fact that the four-wave mixing tensor, for different geometries, can be separated into a small number of parallel factors; the number of such factors is found to scale as the number of modes in the system. The scaling with $M$ can be understood exactly in the case of GRIN fibers (see Appendix titled ``Why is the four-wave-mixing tensor low rank?''). The result of Figure 1 indicates that the argument extends (despite not being analytically extensible) to other geometries, including non-separable index profiles. Additionally, while I have presented examples in optical fibers, the same idea holds for systems where the linear physics involves coupled waveguides (e.g., multi-core fiber), vectorial modes from a numerical eigensolver, or integrated waveguides with low symmetry.

In cases with separability (e.g., $xy$-separability in GRIN, or radial-angular separability in circularly symmetric systems), there is an additional speedup that is achievable beyond what is demonstrated here. To see it, note that the four-wave mixing tensor has a natural CP representation in terms of the mode functions $u$ as $\gamma_{mnpq} \approx \sum_{i} dA_i u^*_m(\boldsymbol{\rho}_i)u^*_n(\boldsymbol{\rho}_i)u_p(\boldsymbol{\rho}_i)u_q(\boldsymbol{\rho}_i)$. Taking the case of circular symmetry as an example, $m = (p,\ell)$ and $u_{p\ell}(r,\phi) = R_{p\ell}(r)e^{i\ell\phi}$ where $R$ is a radial eigenfunction. The contractions of the form $u\cdot A, u^*\cdot A^*$ that appear in $\gamma_{mnpq}A^*_nA_pA_q$ then involve significantly less than $MR \sim M^2$ operations per contraction as the radial and angular parts of the projections can be partially separated. In the large $M$ limit, one expects a $M^{3/2}$ scaling. The special case for circularly symmetric systems has proposed by Laegsgaard \cite{laegsgaard2017efficient,laegsgaard2020multimode}, albeit not implemented at scale. It is the optimal method for high modes with circular symmetry (a commonly considered case). Stated in a tongue-in-cheek way, an improvement exists because the parallel factors have parallel factors. I note that relative to these works: the contribution of the present paper is (1) to show that the compression is general beyond analytically-amenable and symmetric geometries (and can be done purely algebraically on the 4WM tensor itself), (2) to implement it in a public solver and explicitly demonstrate access to ultrafast multimode simulations with very high mode counts, and (3) to characterize performance with other fast implemented methods.

\textit{Results beyond Figures 1-3}. While I have presented GPU timings, the exact same approach leads to very strong speedups on a single-core CPU, and a CPU-compatible CP compression is available in the code accompanying this work \footnote{I have for example been able to simulate the thirty-mode example described in the Supplement over 20k z-steps, in roughly one hour on a 2025 Macbook Pro.}. The speedups are documented in a public code. Although the examples presented here focus on anomalous dispersion, the compression is independent of the dispersion regime in the MMGNLSE, and examples with normal dispersion are in the public code repository. Disorder and polarization are straightforward to include into the CP approach and polarized examples are also available in the public code. A current limitation of the simulations is that the mode-profiles are treated as frequency independent, but it is straightforward to incorporate this without inducing a runtime explosion. 

Typically, it has been the case that because of the aggressive scaling of the MMGNLSE runtime with mode count, that one has to resort to full-field (3+1)D methods which have their own problems. In the Appendix, I show explicitly that the reduction of scaling from quartic to quadratic leads to the modal representation being favorable to a comparable full-field simulation at matched accuracy even at 150 modes, and likely for a much larger number of modes (estimated to be on the order of 1,000). 

\textit{Outlook}. The compressibility of the four-wave-mixing tensor should enable a wider range of exploration in MMNLO enabling multi-parameter sweeps, model-fitting, system design, and explicit optimization to all be significantly more accessible than with the standard dense algorithm, and may enable a broader acceleration of the experiment-theory loop in multimode nonlinear optics. The high-dimensional nonlinear tensor is common to other nonlinear wave systems such as acoustics, water waves, phononics, magnonics, and Bose condensates, and may lead to acceleration in those domains as well.

\section{Acknowledgments}
The author acknowledges start-up funding from the School of Applied and Engineering Physics at Cornell University. The author acknowledges fruitful conversations with Michael Horodynski, Ido Kaminer, Yi-Hao Chen, Myungjoon Kim, and Frank Wise. The author also acknowledges early benchmark data generated by Myungjoon Kim that was used to validate the code. For compute: the author acknowledges the Cornell University Center for Advanced Computing for access to the Seneca High-Performance Computing cluster. AI models were used heavily as coding assistants during the execution time of this project (GPT-5.5, GPT-5.6 Sol, Claude Fable 5, and GPT-6 Astra). The appendix titled "Code Overview" discusses how the code was validated and the examples in the public repository were all written by the author. The author also discussed and developed various mathematical results in collaboration with these models.

\subsection{Code Availability}
The solver developed in this paper, as well as notebooks containing the examples discussed in this paper (and others) can be found in this repository: https://github.com/nrivera494/FastDifferentiableMMNLO. 

\section{Appendices}

\subsection{Why is the four-wave-mixing tensor low rank?}
The persistent compressibility of the 4WM tensor in different geometries, and the approximately observed linear scaling of rank with mode count, points to a more general explanation.  For the case of a GRIN fiber, linear scaling can be analytically shown \footnote{I note that the Gauss-Hermite construction described below was identified by GPT-5.5 Pro and Claude Fable 5. I regard this as providing important conceptual clarity and my contribution in this section is only to spell out the argument and make it digestible.}. The first key observation is that the standard way of computing $\gamma_{mnpq}$ is itself a CP decomposition. In particular, we can write the tensor elements as
\begin{equation}
\gamma_{mnpq} \approx \sum\limits_{i} dA_i u_m(\boldsymbol{\rho}_i)u_n(\boldsymbol{\rho}_i)u_p(\boldsymbol{\rho}_i)u_q(\boldsymbol{\rho}_i).
\end{equation}
where $dA_i$ is an infinitesimal area and $\boldsymbol{\rho}_i = (x_i,y_i)$ are discrete integration points. I have taken real modes for simplicity. If the integration points are sufficient dense, this can be made close to exact. It is also a special case of the CP decomposition of the form Eq.~\eqref{eq:cp-decomposition} where the four vectors are the same (it is a symmetric CP decomposition). The rank of this CP decomposition is the number of integration points. It is worth emphasizing that while the standard way of writing the overlap is a CP decomposition, the more general four-factor form in Eq.~\eqref{eq:cp-decomposition} is the parameterization we employ and it achieves a rank far below what is expected from a naive Cartesian choice of grid points. For the GRIN case, the use of a uniform Cartesian grid of integration points and uniform integration weights is extremely suboptimal. In particular, the overlaps can be represented by a small number of Gauss-Hermite quadrature points.

The elements of the 4WM tensor for a GRIN fiber, in a Cartesian basis (with Hermite-Gauss eigenfunctions proportional to $e^{-\frac{1}{2}(x^2+y^2)}H_{m_x}(x)H_{m_y}(y)$ in suitably nondimensionalized coordinates), are schematically of the form $\gamma_{mnpq} = I_x I_y$ with 
\begin{widetext}
\begin{equation}
I_x \sim \int dx ~e^{-2x^2}H_{m_x}(x)H_{n_x}(x)H_{p_x}(x)H_{q_x}(x) \approx \sum\limits_{q=1}^{P} w_q   H_{m_x}(x_q)H_{n_x}(x_q)H_{p_x}(x_q)H_{q_x}(x_q).
\end{equation}
\end{widetext}
This integral is precisely the type that Gauss-Hermite quadrature is meant to approximate. The maximum monomial order that can appear in the integral above, assuming we retain $G$ degenerate mode groups of the GRIN fiber (corresponding to a total number of modes $M = G(G+1)/2$) is $4(G-1)$. This can be represented exactly using $2G$ quadrature points.  The two-dimensional overlap can therefore be exactly integrated by $4G^2 \approx 8M$ points. This coincides with the CP rank for \emph{exact} representation in the GRIN case. A good CP decomposition with finite tensor error $\epsilon$ should then achieve a rank below $8M$. 

Going back to the finding of Figure 1a for step-index and asymmetric fiber, we can think that the tensor has some ideal quadrature representation albeit one that isn't analytically obvious unlike the GRIN case. In other words, the hypothesis is that the low-rank nature really just depends on some kind of smoothness not specific to GRIN. The fact that the CP decomposition finds similar ranks and approximate linear scaling for other fibers is an indication that these other geometries also have low-rank structure. Despite these other fibers not having an obvious exact construction, the CP decomposition generically and algorithmically finds the low-rank representation.

It is worth pointing that this argument is very closely related to an approach described by Laegsgaard to simplify modal simulations in GRIN and step-index fibers, despite being described in slightly different language \cite{laegsgaard2017efficient,laegsgaard2020multimode}. It was found in their work that by evaluating the 4WM matrix elements on a sparse quadrature grid (e.g., with Gauss-Legendre quadrature points) that the complexity of evaluating the nonlinear polarization (the same quantity up to prefactors as the nonlinear contraction) scales like $MP$ with $P \sim M$ being the number of quadrature points. An explicit choice of points was found for GRIN and other circularly symmetric fibers such as step index. Therefore, the statements (1) there is an efficient quadrature representation of the nonlinear polarization, and (2) the 4WM tensor has a low-rank representation, are the same statement. I note that relative to these works: the contribution of this paper is (1) to show that the compression is general beyond analytically-amenable and symmetric geometries (and can be done purely algebraically on the 4WM tensor itself), (2) to implement it in a solver and explicitly demonstrate access to ultrafast multimode simulations with very high mode counts, and (3) to characterize performance.

\subsection{Comparison to full-field representation}

The more favorable scaling of the compressed approach is also important to consider when deciding to run a simulation in the modal representation or in a full-field, 3+1D representation. The original spacetime ("full-field") representation of the nonlinear Schrodinger equation $A(\boldsymbol{\rho},z,t)$ is given by (in a weak guidance approximation):
\begin{widetext}
\begin{equation}
    \partial_z A = \sum\limits_{k=0}^{\infty} \frac{i^{k+1}\beta_{k}}{k!} \partial_t^k A + i\left[\frac{1}{2\beta_0}\nabla^2_{\perp} + \frac{\omega_0}{c}\Delta n(\boldsymbol{\rho}) \right]A + i\frac{n_2\omega_0}{c}\left(1 + i\tau_S\partial_t \right)\left((1-f_R)|A|^2A + f_R (h_R \ast |A|^2) A\right)
\end{equation}
\end{widetext}
Importantly, the nonlinear term is diagonal in real-space and is not subject to the quartic complexity associated with the modal representation. As a reminder, the MMGNLSE makes an implicit assumption that the electric field can be expressed as a sum over waveguide modes, and therefore cannot be expected as-is to resolve effects related to radiation loss or linear/nonlinear couplings to modes outside of the retained subspace. Supposing that we are interested in simulating a system that stays in our initially prescribed modal subspace to start, why bother with modal representations? There are a few reasons. For one, the linear physics is treated exactly: the modal representation works with the exact mode profiles and importantly propagation constants, if the modes are analytical (but a numerical mode solver with sufficient discretization can also get the modes accurately). In the full-field approach, one works with the modes prescribed by the transverse Laplacian plus index on a spatial discretization. In principle one can work with a spatial discretization that matches that of a mode solver, but in practice that would imply an extremely large state dimension and an expensive propagation step. Over a meter of propagation, extremely small wavevector-detuning errors of even $1 m^{-1}$ are large, and so inaccuracies in the linear physics get magnified quickly. In other words: if the physics is modal, a modal simulation is often preferable. It has also been noted that the modal approach more readily yields conceptual insight \cite{wright2017multimode}. 

\begin{figure}[t]
    \centering
    \includegraphics[width=0.45\textwidth]{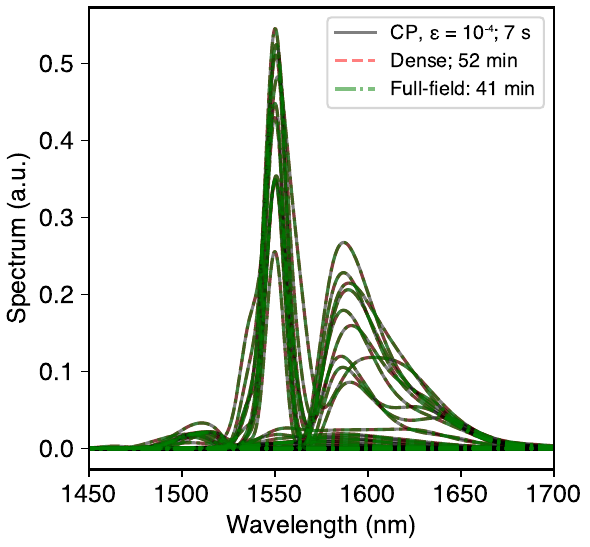}
    \caption{\textbf{Benchmarking CP decomposition against full-field and dense modal solvers.} Modal spectra predicted by a modal and full-field representation, designed such that the relative norm error in the modal spectra is roughly $5 \times 10^{-4}$ between CP and dense and between full-field and dense. 150 modes are retained in the simulation, which involves an initial ten-mode (modes 6-15) launch in step-index fiber propagating through 0.25 meters of fiber.}
\end{figure}

\begin{table*}[t]
\centering
\renewcommand{\arraystretch}{1.55}
\setlength{\tabcolsep}{3.5pt}
\newcommand{\exampletableheader}[2]{%
  \parbox[c][8.2ex][c]{1.75cm}{\centering\footnotesize #1\\[-0.1ex]
    {\scriptsize\color{black!65}#2}}%
}
\newcommand{\examplenotebook}[2]{%
  \href{https://github.com/nrivera494/FastDifferentiableMMNLO/blob/main/#1}{#2}%
}
\resizebox{\textwidth}{!}{%
\begin{tabular}{|l|c|c|c|c|c|c|c|c|c|c|}
\hline
\multicolumn{1}{|c|}{\textbf{Capability / feature}}
 & \exampletableheader{\examplenotebook{example_notebooks/squeezing_spectral_filter.ipynb}{Raman soliton\\filtering}}{}
 & \exampletableheader{\examplenotebook{example_notebooks/soliton_jitter.ipynb}{Soliton\\jitter}}{}
 & \exampletableheader{\examplenotebook{example_notebooks/step_index_squeezing.ipynb}{Step-index\\squeezing}}{}
 & \exampletableheader{\examplenotebook{example_notebooks/om4_1030.ipynb}{Normal-\\dispersion CP}}{}
 & \exampletableheader{\examplenotebook{example_notebooks/birefringent_mm_soliton.ipynb}{Birefringent\\solitons}}{}
 & \exampletableheader{\examplenotebook{example_notebooks/gpu_examples/stochastic_example_gpu.ipynb}{Stochastic\\fission}}{}
 & \exampletableheader{\examplenotebook{example_notebooks/gpu_examples/cp_decomp_modecount_gpu.ipynb}{CP mode-count\\scaling}}{}
 & \exampletableheader{\examplenotebook{example_notebooks/gpu_examples/full_field_vs_cp_gpu.ipynb}{Full-field\\vs. CP}}{}
 & \exampletableheader{\examplenotebook{example_notebooks/gpu_examples/manymode_prop_gpu.ipynb}{1,035-mode\\propagation}}{}
 & \exampletableheader{\examplenotebook{example_notebooks/gpu_examples/cp_decomposition_30mode_soliton_gpu.ipynb}{30-mode CP\\benchmark}}{} \\
\hline
Dense forward      & \cellcolor{blue!50} & \cellcolor{blue!50} & \cellcolor{blue!50} & \cellcolor{blue!50} & \cellcolor{blue!50} &                       & \cellcolor{blue!50} & \cellcolor{blue!50} &                       & \cellcolor{blue!50} \\
\hline
CP forward         &                       &                       & \cellcolor{blue!50} & \cellcolor{blue!50} &                       & \cellcolor{blue!50} & \cellcolor{blue!50} & \cellcolor{blue!50} & \cellcolor{blue!50} & \cellcolor{blue!50} \\
\hline
Full-field forward &                       &                       &                       &                       &                       &                       &                       & \cellcolor{blue!50} &                       &                       \\
\hline
Dense adjoint      & \cellcolor{blue!50} & \cellcolor{blue!50} &                       &                       & \cellcolor{blue!50} &                       &                       &                       &                       &                       \\
\hline
CP adjoint         &                       &                       & \cellcolor{blue!50} &                       &                       &                       &                       &                       &                       &                       \\
\hline
Stochastic         &                       &                       & \cellcolor{blue!50} &                       &                       & \cellcolor{blue!50} &                       &                       &                       &                       \\
\hline
Distributed noise  & \cellcolor{blue!50} & \cellcolor{blue!50} &                       &                       &                       &                       &                       &                       &                       &                       \\
\hline
Time               & \cellcolor{blue!50} & \cellcolor{blue!50} & \cellcolor{blue!50} & \cellcolor{blue!50} & \cellcolor{blue!50} & \cellcolor{blue!50} & \cellcolor{blue!50} & \cellcolor{blue!50} & \cellcolor{blue!50} & \cellcolor{blue!50} \\
\hline
Spatial modes      &                       &                       & \cellcolor{blue!50} & \cellcolor{blue!50} & \cellcolor{blue!50} & \cellcolor{blue!50} & \cellcolor{blue!50} & \cellcolor{blue!50} & \cellcolor{blue!50} & \cellcolor{blue!50} \\
\hline
Polarization       &                       &                       &                       &                       & \cellcolor{blue!50} &                       &                       &                       &                       &                       \\
\hline
GPU                &                       &                       &                       &                       &                       & \cellcolor{blue!50} & \cellcolor{blue!50} & \cellcolor{blue!50} & \cellcolor{blue!50} & \cellcolor{blue!50} \\
\hline
External validity  &                       & \cellcolor{blue!50} &                       &                       & \cellcolor{blue!50} &                       &                       &                       &                       & \cellcolor{blue!50} \\
\hline
Internal validity  & \cellcolor{blue!50} & \cellcolor{blue!50} & \cellcolor{blue!50} & \cellcolor{blue!50} & \cellcolor{blue!50} &                       & \cellcolor{blue!50} & \cellcolor{blue!50} & \cellcolor{blue!50} & \cellcolor{blue!50} \\
\hline
\end{tabular}%
}
\vspace{0.4em}

\parbox{0.98\textwidth}{\footnotesize
\textcolor{blue!50}{\rule{1.25em}{1.0em}} corresponds to a capability explicitly demonstrated in the corresponding example. The examples correspond to notebooks in \texttt{example\_notebooks/} in the GitHub repository. GPU columns correspond to notebooks in \texttt{example\_notebooks/gpu\_examples/}. External validity and internal validity are explained in the text. Columns link to example notebooks in the text.}
\end{table*}

At the same time, even in cases where the dynamics mostly live in a finite modal subspace, the aggressive scaling of the modal representation often yields a mode-number for which a full-field simulation becomes cheaper. Typically, other works have reported this crossover to be for five to ten modes \cite{eslami2022two}. Here, I compare full-field and modal representations of the same dynamics. The conclusion is that even at 150 modes, the modal representation has a large, decisive advantage at matched accuracy. Figure 4 shows the spectra of each mode after an initial ten-mode soliton launch (modes six through fifteen) of a step-index fiber of 52.5 micron core radius and numerical aperture of 0.275 \footnote{The original file used to generate the data is available, hence I do not report every parameter here.}, calculated three ways: (1) with a full-field simulation of the (3+1)D nonlinear Schrodinger equation using a second-order integrator \footnote{A fourth-order integrator was found to be slower than second-order, even at larger step-size, presumably due to the complexity of the fourth-order update step.}, (2) with a dense reference simulation at a step size of 200 microns, and (3) with a CP decomposed nonlinear tensor with a tensor error of $\epsilon = 10^{-4}$. The two modal simulations use a fourth-order integrator based on a standard RK4IP approach (fourth-order Runge-Kutta on the interaction picture MMGNLSE transforming away the linear terms). Importantly, the modes and propagation constants used in the modal representation are determined by the eigenvectors and eigenvalues of the transverse Laplacian and the potential set by the index of refraction defined on a 64-by-64 transverse grid, in order to make sure the linear physics is the same in all three simulations. We compare the CP and full-field at "matched accuracy": in other words, we find the time needed to get a solution whose error (norm error) relative to the dense simulation is roughly the same ($5 \times 10^{-4}$ for this example). The dense is taken as the reference since it is seen that reducing the step size further on the full-field approach makes the solution more closely approach the dense solution. The full-field simulation needs a significantly smaller step to achieve the same accuracy (1.25 microns), while the modal approach is already sufficiently converged at a much larger step (200 microns) $-$ a fourth-order integrator for the full-field enables a larger step but still significantly below 100 microns. The comparison is striking: at 150 modes, the dense approach takes nearly an hour, the full-field approach roughly 40 minutes, and the CP approach roughly ten seconds \footnote{150 was chosen as a clean number of modes that is still somewhat cheaply decomposable, and where a nontrivial dense simulation can be executed on the H100 SXM. Importantly, the decomposition itself takes roughly ten minutes on the same H100 SXM GPU. I do not include this in the speedup because the best use-case is when the same fiber is simulated many times, in which case the cost is greatly diluted.}.

\subsection{Code overview}
A GPU-compatible MMGNLSE solver which uses a CP-decomposed four-wave-mixing tensor is implemented in public code (FastDifferentiableMMNLO.jl) and is part of a larger set of routines.  An overview of the most important code capabilities are as follows:
\begin{enumerate}
\item CPU and GPU implementations of the multimode generalized nonlinear Schrodinger equation (with dispersion, linear gain/loss, self-steepening, Raman, and polarization physics included). There is a standard ``dense'' (quartic-mode-scaling) implementation and an implementation that takes and uses the compressed tensor. The solver routines include fixed-step interaction-picture methods using RK4 (fourth-order Runge-Kutta) and Tsit5 (fifth-order Tsitouras). The solver also supports in some cases adaptive-stepping and mixed-precision.
\item GPU implementation of a full-field (3+1) solver using second-order Strang splitting of the linear (kinetic+potential) part of the equation.
\item A stochastic implementation that allows one to specify an ensemble of initial conditions that are then propagated through the MMGNLSE.
\item Adjoint implementations of the modal MMGNLSE (CPU and GPU implementation, as well as dense and CP implementations), which allow one to compute the gradient of an arbitrary observable with respect to the initial conditions of the MMGNLSE. This is useful for optimization, inverse design, and quantum noise effects in the linearized quantum noise approximation \cite{zia2025noise,sloan2025programmable}.
\end{enumerate}
The available code features a variety of examples which exercise these different features of the code and also compare its outputs to existing codes in cases where there is overlap.

I include a table here which illustrates the examples in the repository, and the notebook. Some of the example notebooks in the repository make these comparisons directly so that you can look at them for yourself. Most run on a single CPU in a reasonable amount of time. The table also describes measures of internal and external validity. Internal validity means I compared multiple ways to do the calculation (e.g., comparing an adjoint gradient against reverse-mode automatic differentiation, or I compared modal representation against full-field for a forward solver). It can also mean that limiting cases were checked (e.g., photon-number conservation). External validity means: we checked what our code says against what someone else's code says, or what an analytical prediction says, and they say the same thing. 

\bibliographystyle{unsrt}
\bibliography{qnoise.bib}

\end{document}

%% file: preamble.tex
\usepackage{microtype} 
\usepackage{xspace} 

\usepackage{xr}

\usepackage{xcolor}
\usepackage{graphicx}
\usepackage{tikz}
\usetikzlibrary{calc,decorations.markings}

\usepackage{hyperref}
\hypersetup{
    colorlinks,
    linkcolor={blue!75!black!80!yellow},
    citecolor={blue!75!black!80!yellow},
    urlcolor={blue!75!black!80!yellow}
}

\usepackage{empheq}

\let\oldmarginpar\marginpar 
\def\marginpar#1{\oldmarginpar{\scriptsize #1}}
\usepackage[capitalize,nameinlink]{cleveref}

\crefname{subequations}{Eqs.}{Eqs.} 
\Crefname{subequations}{Eqs.}{Eqs.}
\crefformat{subequations}{#2Eqs.~(#1)#3}
\Crefformat{subequations}{#2Eqs.~(#1)#3}
\crefname{page}{p.}{p.} 
\newcommand{\heqref}[1]{\hyperref[#1]{(\ref*{#1})}} 
\newcommand{\namedref}[2]{\hyperref[#2]{#1~\ref*{#2}}} 
\newcommand{\namedeqref}[2]{\hyperref[#2]{#1~(\ref*{#2})}} 

\usepackage[toc,page]{appendix}

\usepackage{enumitem}

\newcommand{\appropto}{\mathrel{\vcenter{
  \offinterlineskip\halign{\hfil$##$\cr
    \propto\cr\noalign{\kern.2pt}\sim\cr\noalign{\kern-2.5pt}}}}}